# Phase characteristics of electromagnetically induced transparency analogue in coupled resonant systems

**Xiaoyan Zhou[1], Lin Zhang[2], Wei Pang[1], Hao Zhang[1], Qingrui Yang[1] and Daihua Zhang[1]**

[1] State Key Laboratory of Precision Measuring Technology and Instruments,
Tianjin University, Tianjin 300072, China.

[2] Microphotonics Center and Department of Materials Science and Engineering,
Massachusetts Institute of Technology, Cambridge, MA 02139, USA.

Email: linzhang@mit.edu and weipang@tju.edu.cn

**Abstract.** Electromagnetically induced transparency (EIT) and EIT-like effects have been investigated in a wide variety of coupled resonant systems. Here, a classification of the phase characteristics of the EIT-like spectral responses is presented. Newly identified phase responses reveal unexplored operation regimes of EIT-like systems. Taking advantage of the new phase regimes, one can obtain group delay, dispersion and nonlinearity properties greatly enhanced by almost one order of magnitude, compared to the traditionally constructed EIT-like devices, which breaks the fundamental limitation (e.g., delay-bandwidth product) intrinsic to atomic EIT and EIT-like effects. Optical devices and electrical circuits are analyzed as examples showing the universality of our finding. We show that cavity-QED-based quantum phase gates can be greatly improved to achieve a phase shift of $\pi$. The new phase characteristics are also believed to be useful to build novel doubly resonant devices in quantum information based cavity QED, optomechanics, and metamaterials.

**1. Introduction**

Electromagnetically induced transparency (EIT) has been intensely investigated in the past decades [1, 2]. Extensive research efforts have been made in fundamental physics and exciting applications. These include quantum information [3-5], lasing without inversion [6], optical delay [7, 8] (sometimes called "slow light" [9]), nonlinearity enhancement [10, 11], precise spectroscopy [12], pushing frontiers in quantum mechanics and photonics. EIT was first observed in atomic media [1, 2]. Featuring a nearly transparent window in an "absorption" spectrum, EIT-like effects are identified as a universal phenomenon in coupled resonant systems, as shown in figure 1, in optics [13-16], mechanics and electrical circuit [17], plasmonics and metamaterials [18-21], and hybrid configurations [22-26]. In these coupled resonant systems, the basic physical principle underlying is the interference of fields instead of probability amplitudes as in a three-level atomic system [14]. This renders EIT-like systems flexible and controllable candidates for diverse functional devices. Although, in physics, amplitude and phase are generally viewed as equally important quantities, one tends to identify the EIT-like effects from various new materials and structures only by reporting an EIT-like intensity response, and the phase characteristics of EIT-like effects has long been remaining unclear, compared to the striking signature in intensity spectra.

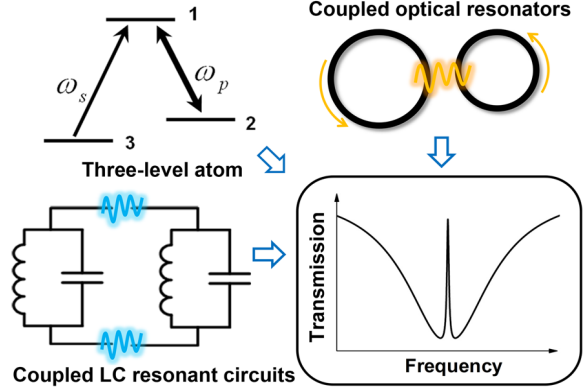

**Figure 1.** Various coupled resonant systems exhibiting EIT or EIT-like transmission spectrum.

In general, a physical resonant system can be characterized by its damping and driving (sometimes called coupling in optics). The relationship between the two factors determines the "polarity" of the resonance: here a resonator with coupling stronger or weaker than damping is considered "positive polarity" or "negative polarity", respectively. It is important to note that the response of the resonator to an input stimulus is strongly dependent on its "polarity". A good example is that, in optical domain, a resonator is typically differentiated to be under-coupled, critically coupled, or over-coupled [27]. As another example, in cavity QED, strong coupling of a photon in the cavity mode with an intra-cavity atom or quantum dot is a preferred operation regime [28, 29]. Not only as a physical concept, identifying the "polarity" of a resonator and constructing the desired "polarity" of a resonator-based device are essential for achieving many functional components in both classical physics [30-33] and quantum physics [34-36]. Since the EIT-like effects are generally found in multi-resonant systems, it would be critically important to analyze and tailor the "polarity" of each resonant element.

In this paper, we reveal the existence of other phase regimes in EIT-like effects, besides the well-known phase anomaly in atomic EIT, under the same EIT-like intensity response. We classify them in terms of cavity "polarity". New phase responses represent an unexplored aspect of EIT-like systems. One can thus enhance dispersion and nonlinearity by almost one order of magnitude compared to previous EIT-like devices, breaking the fundamental limitation (e.g., delay-bandwidth product) intrinsic to atomic EIT and EIT-like effects. To show the universality of new phase responses and the generality of their categorization, we analyze optical devices and electrical circuits as examples. We also briefly discuss the impact of the new phase regimes on important physics branches (quantum information based cavity QED, optomechanics, and metamaterials), with an emphasis on the enhancement of quantum phase gates based on cavity quantum electrodynamics (QED). More than twice improvement of phase shift, as large as $\pi$, is obtained in the newly indentified phase regime, which is highly desirable for basic quantum logic gates in quantum computation.

## 2. Phase regimes of EIT-like effects in coupled optical resonators

### 2.1. Four types of phases with the same EIT intensity profile

First, we consider a coupled resonator system consisting of two identical optical microring resonators between two waveguides in figure 2. The resonators are coupled to the waveguides with power coupling coefficients $c_1$, $c_2$, $c_3$ and $c_4$. Although reported [15], this device is only designed to have coupling coefficients that are all equal or very close to each other. Due to cavity loss, the two resonators are under-coupled in reality. Here, we break the symmetry of the coupling between the rings and the waveguides. For simplicity, the coupling between the resonators is ignored. As an example, we consider $Si_3N_4$ cavities with the azimuthal mode orders of the resonators are $m_1 = m_2 = 129$. The straight waveguides between the resonators have an equivalent mode order $m_3 = 47$. The transfer function, $E_{out}(\omega)/E_{in}(\omega)$, of this system is derived using coupled mode theory (CMT) [37], where $E_{in}(\omega)$ and $E_{out}(\omega)$ are optical fields at the "In" and "Out" ports in figure 2. Transmission and phase are defined as $|E_{out}(\omega)/E_{in}(\omega)|^2$ and $angle(E_{out}(\omega)/E_{in}(\omega))$, respectively.

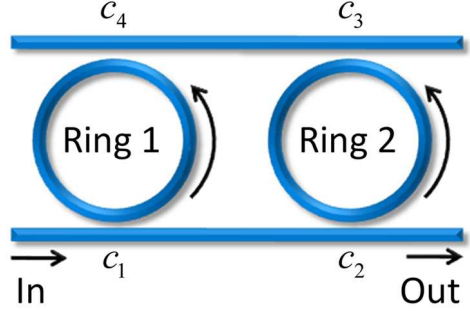

**Figure 2.** Schematic of the optical coupled resonators.

For simplicity without losing generality, we first ignore optical loss, and thus the "polarity" of each cavity is determined by the relationship between in-coupling ($c_1$, $c_2$, i.e., the driving) and out-coupling ($c_3$, $c_4$, i.e., the damping). In figure 2, $c_1$ and $c_4$ are chosen to be 0.008 and 0.02, while $c_2$ and $c_3$ are set to be 0.01 and 0.04, respectively. By switching the coupling coefficients of Ring 1 and Ring 2, respectively, we identify four types of phase responses, and they all have the same EIT-like intensity response, as shown in figure 3 (solid lines in the transmission and phase profiles). With $c_1 = 0.008$, $c_4 = 0.02$, $c_2 = 0.01$, and $c_3 = 0.04$, both resonators are under-coupled, and the phase profile in figure 3(a) features a non-monotonic shift across the resonance frequency. This was reported together with the EIT-like intensity response [15]. Such a phase anomaly corresponds to the well-known phase characteristics in atomic EIT phenomena [38], and we name it Type I. When we switch $c_2$ and $c_3$, Ring 2 is changed to the over-coupled regime, with Ring 1 still under-coupled. The exactly same EIT-like intensity response is obtained, but, intriguingly, the phase response is dramatically changed, being monotonically increasing with frequency from 0 to $2\pi$, as shown in figure 3(b). A similar phase profile with the same intensity response in figure 3(c) can also be observed, when Ring 2 remains under-coupled and Ring 1 is turned to be over-coupled by switching $c_1$ and $c_4$. These two $2\pi$ phase profiles in figures 3(b) and 3(c) correspond to different "polarity" states of the coupled resonators, which causes a different slope of phase change, and we name them Types II and III. Finally, by setting both resonators over-coupled, we obtain a phase shift of $4\pi$ in figure 3(d), which is called Type IV. Note that the phase characteristics in the EIT-like effects, called Types II, III, and IV, have not been investigated in literature.

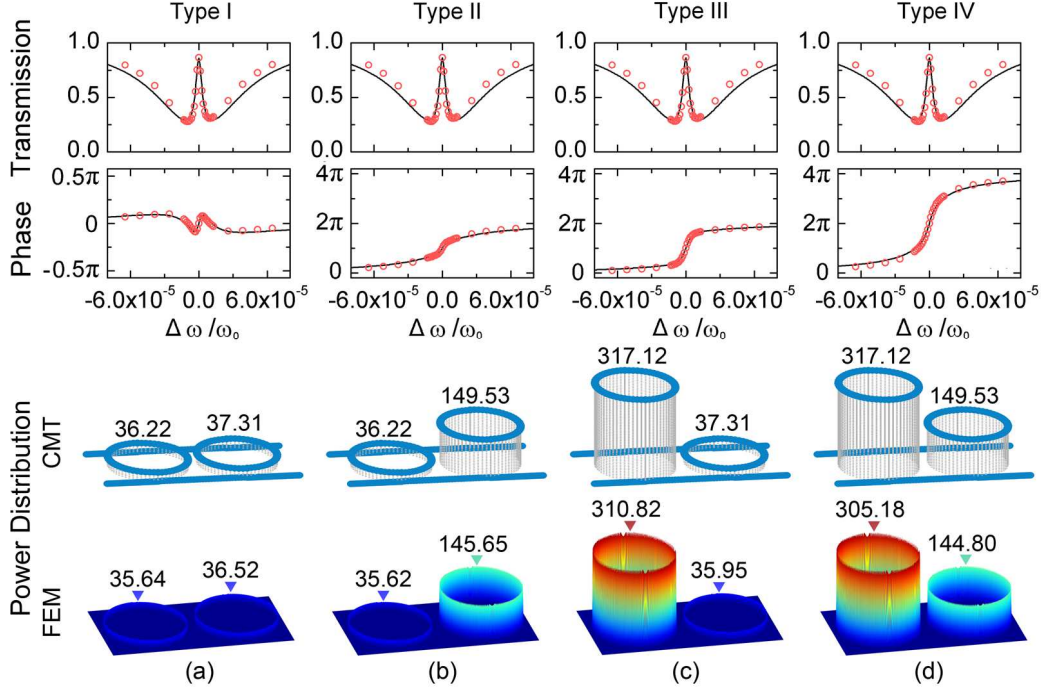

**Figure 3.** Transmission and phase spectra of the EIT-like effect in coupled optical resonators and corresponding power distributions for (a) Type I, (b) Type II, (c) Type III, and (d) Type IV, respectively. Transfer characteristics are obtained with both coupled mode theory (CMT), denoted by solid lines, and finite element method (FEM), denoted by red circles. Power distributions are also obtained with both CMT and FEM.

Also in figure 3, we show the normalized power distributions of the four EIT types, corresponding to the phase profiles above, with loss ignored. At the resonance frequency, the accumulated optical power in the cavities greatly varies from type to type. To evaluate the nonlinearity enhancement in the presented coupled-resonant system, one can treat it as a nonlinear medium and calculate the nonlinear refraction (e.g., we consider 3$^{rd}$-order Kerr nonlinearity here) [39]: $\Delta n = n_2 P/A_{eff}$, where $P$ is the optical power, $A_{eff}$ is the effective mode area, and $n_2$ is the Kerr nonlinear index related to $\chi^{(3)}$ by $n_2 = 3/(4\varepsilon_0 cn)\cdot\text{Re}(\chi^{(3)})$ [40]. This way, the nonlinear refraction $\Delta n = 3\text{Re}(\chi^{(3)})P/(4\varepsilon_0 cnA_{eff})$, which is proportional to the product of $P$ and $\text{Re}(\chi^{(3)})$. Therefore, the EIT-like resonance enhancement to the input optical power in our case is equivalent to $\chi^{(3)}$ enhancement in atomic EIT media [2], to obtain the same the nonlinear refraction $\Delta n$. Large power enhancement, i.e., strong $\chi^{(3)}$ nonlinearity enhancement, is obtained in a certain ring, when it is turned to be over-coupled. In Type I and II, $\chi^{(3)}$ nonlinearity enhancements are 37.31 and 149.53, respectively; while large $\chi^{(3)}$ enhancement of 317.12 is obtained in Type III and Type IV, which is 8.5 times that in Type I, and 2.1 times that in Type II. In addition, we note that the power enhancement factors in Ring 1 are the same for Type I and Type II in figures 3(a) and 3(b). This is because Ring 1 has its "polarity" unchanged in the two types. A similar trend is also found with Ring 2, as shown in figures 3(a) and 3(c). That is, the power distribution in one of the two coupled resonators is dominantly determined by its own "polarity", with negligible influence from the other resonator, although optical fields in the two resonators strongly interact when the EIT-like effect occurs.

Numerical simulation of the coupled resonator system in figure 2 is conducted using a finite-element-method (FEM) solver, which verifies the results obtained with CMT. In order to have the same parameters as above, we set the waveguide width to be 330 nm. With azimuthal mode orders $m_1 = m_2 = 129$, the ring resonators have a radius of 20 μm, and the resonance wavelength is calculated to be 1553.9 nm. To realize the above coupling coefficients: 0.008, 0.01, 0.02, and 0.04, we locally taper the widths of the straight waveguides in the coupling regions to be 430, 415, 366, and 284 nm, respectively. Note that the distance between central lines of the ring-shaped waveguides and straight waveguides is 1 μm. In excellent agreement with the analytical results by CMT, four types of phase responses with almost unchanged EIT-like transmission are observed, as shown by red circles in figure 3. We also show power distributions from FEM simulations for the four phase regimes, which are almost the same as those from CMT.

*2.2. Physical origin of the newly identified EIT phase regimes*

To gain a better understanding of the phase characteristics in the EIT-like effects, we use the dynamic CMT [41] to describe the coupled resonator system. The straight waveguides are assumed to have frequency-independent phase shifts near the resonance and are set to be lossless, since optical power is mainly confined in the cavities. The dynamic equations describing the evolution of energy in the resonators are:

$$\frac{da_1}{dt} = (-i\omega_1 - \kappa_1 - \kappa_4 - \gamma_1) \cdot a_1 + i\sqrt{2\kappa_1} E_{in} + i\sqrt{2\kappa_4} \cdot i\sqrt{2\kappa_3} a_2 \quad (1)$$

$$\frac{da_2}{dt} = (-i\omega_2 - \kappa_2 - \kappa_3 - \gamma_2) \cdot a_2 + i\sqrt{2\kappa_2} E_{in} + i\sqrt{2\kappa_2} \cdot i\sqrt{2\kappa_1} a_1 \quad (2)$$

The output electric field is expressed as:

$$E_{out} = E_{in} + i\sqrt{2\kappa_1} a_1 + i\sqrt{2\kappa_2} a_2 \quad (3)$$

where $E_{in}$ and $E_{out}$ are the electric fields at the "In" and "Out" port; $a_j$ ($j = 1, 2$) is the energy amplitude in the $j$th ring and is defined as $a_j = |a_j|e^{-i\omega t}$; $\kappa_k$ ($k = 1, 2, 3, 4$) is the cavity decay rate due to coupling, which is related to $c_k$ by $\kappa_k = c_k v_g/(4\pi R)$; $\gamma_j$ and $\omega_j$ ($j = 1, 2$) are the cavity decay rate due to cavity loss and resonance angular frequency of the $j$th resonator, respectively. The two resonators have the same resonance frequency $\omega_1 = \omega_2$. Solving these equations, we obtain the transfer function for the system:

$$T = \frac{E_{out}}{E_{in}} = \frac{A_1 B_1}{A_2 B_2 - 4\sqrt{\kappa_1 \kappa_2 \kappa_3 \kappa_4}} \quad (4)$$

where $A_j$ and $B_j$ ($j = 1, 2$), related to Ring 1 and Ring 2, are defined as $A_j = i(\omega-\omega_1) - (-1)^j \kappa_1 - \kappa_4 - \gamma_1$ and $B_j = i(\omega-\omega_2) - (-1)^j \kappa_2 - \kappa_3 - \gamma_2$, respectively.

From equation (4), we note that the denominator remains the same, when the coupling factors are switched, and the overall spectral phase of the coupled resonator system is changed only because of the change in the positions of the zeros of the transfer function. With loss neglected (i.e., $\gamma_1 = \gamma_2 = 0$), the two zeros, $Z_1 = \kappa_1 - \kappa_4$ and $Z_2 = \kappa_2 - \kappa_3$, are determined purely by the difference between the cavity decay rates due to coupling. For Type I, where $\kappa_1 < \kappa_4$ and $\kappa_2 < \kappa_3$, $Z_1$ and $Z_2$ are both on the lower half complex frequency plane, and the coupled resonators are working as a minimum-phase system [42]. If the "polarity" of one ring is switched, one of the zeros is shifted to the upper half plane, and the phase response of the system is increased, as shown in figures 3(b) and 3(c). When both rings are over-coupled and $Z_1$ and $Z_2$ are on the upper half plane, the system works in the maximum-phase regime, with a maximum phase shift of $4\pi$ in figure 3(d). In this process, the spectral amplitude of the EIT-like transfer function does not change. Clearly, the expression of the transfer function derived from the dynamic coupled mode theory provides a mathematical proof of the existence of the presented phase characteristics and their relationship to the "polarity" of the cavities, which is also physically intuitive.

Intriguingly, it seems that only Type I phase satisfies the Kramers-Kronig relations, with the EIT-like amplitude response. In fact, the Kramers-Kronig relations hold for the real and imaginary parts of the transfer function of our casual coupled resonance system, which is explained in [43]. Our simulation also confirms this by showing the real and imaginary parts over frequency in figure 4. However, the amplitude and phase responses of the system are not governed by the Kramers-Kronig relations [43], except for the minimum-phase operation regime (i.e., Type I). In Type I, the zeros of the transfer functions, i.e., $Z_1$ and $Z_2$, are both negative, making the natural logarithm of equation 4 analytic in the upper half of the complex frequency plane and the amplitude and phase of the transfer functions an Hilbert transform pair [43]. Structure-induced phase and dispersion are actually designable, and many non-minimum-phase devices are broadly used in various applications [42, 44].

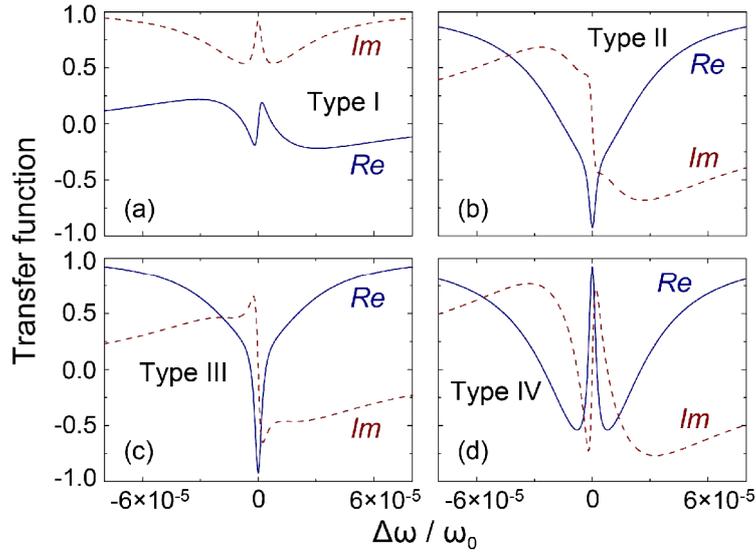

**Figure 4.** (a), (b), (c), and (d) are real (solid line) and imaginary (dashed line) parts of the EIT-like complex transfer functions for the coupled resonators in Types I, II, III, and IV, respectively.

*2.3. Enhanced group delay and nonlinearity in the new EIT phase regimes*

A highly interesting feature of EIT and EIT-like effects is the great reduction of group velocity of light [7, 15, 16], sometimes called "slow" light. Although different physical systems and devices have been demonstrated for "slowing" light, the group delay is always produced via a phase anomaly (i.e., Type I phase in this phase) and enhanced through narrowing of the linewidth of EIT-like transmission. There is a well-known trade-off between group delay and bandwidth, which is intrinsic to atomic EIT and EIT-like effect with a phase anomaly. Figure 5(a) shows the group delays ($\tau_d$) normalized to the cavity round-trip time ($\tau_c$) versus frequency detuning, with all the parameters as in figure 5. The group delay is obtained using CMT and is defined as $\tau_d = \partial(angle(E_{out}(\omega)/E_{in}(\omega))) / \partial\omega$. With the same intensity response, Type IV produces the largest delay on the resonance, which is almost four times that given by the commonly used Type I phase. Moreover, Type IV exhibits the largest bandwidth of delay. Therefore, the newly identified phase regimes, particularly for Type IV, break the fundamental limitation set by the delay-bandwidth product in the traditional EIT-like effects. There are negative or zero group delay regions in Type I, which seems interesting that fast light might be generated. But these regions correspond to the EIT valleys in the amplitude response, where optical loss is high, and therefore are hard to be utilized for practical applications.

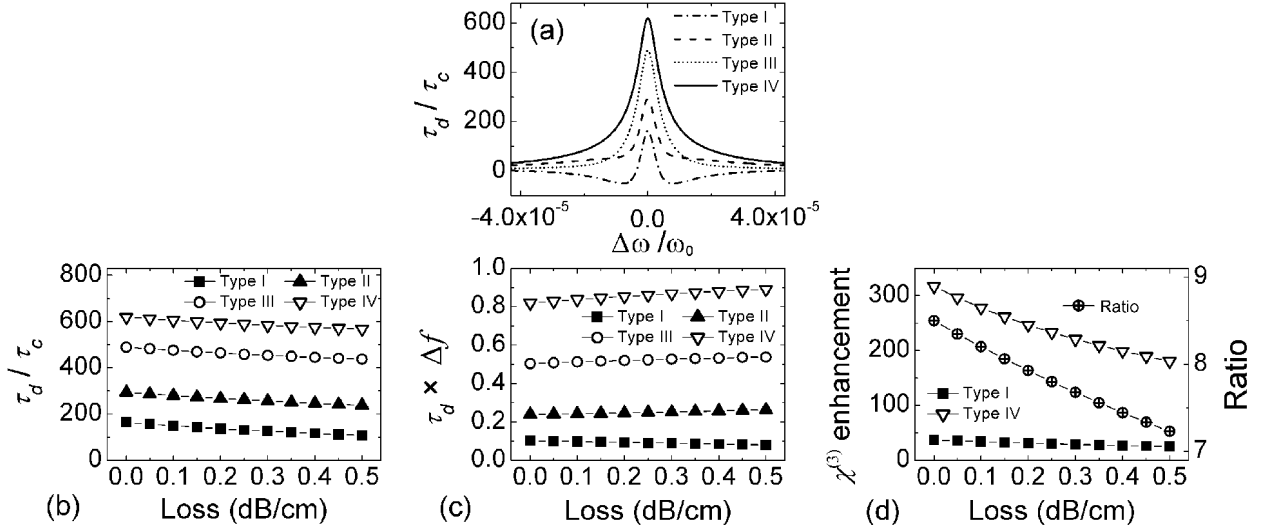

**Figure 5.** (a) Normalized group delay in four types of EIT-like systems. The role of cavity loss on (b) group delay, (c) delay-bandwidth product, and (d) nonlinearity enhancement, respectively.

With optical loss taken into account, the group delay for all the four types decreases with loss, as shown in figure 5(b). The decrement rate is almost the same, i.e., Type IV becomes more advantageous as a delay element, compared to other types. For example, as the loss increases from 0 to 0.5 dB/cm, the group delay in Type IV increases from 3.8 to 5.3 times that in Type I. An interesting phenomenon is found in figure 5(c): The four types of EIT-like effects have different trends in the delay-bandwidth product ($\tau_d \times \Delta f$) as the loss changes, where $\Delta f$ is defined as the full width at half maximum of the normalized group delay. The commonly addressed phase, Type I, has a delay-bandwidth product decreasing with loss. When the resonant system becomes more over-coupled transitioning from Type I to Type IV, the delay-bandwidth product can even increase with loss [see, e.g., Types II, III and IV in figure 5(c)]. With a loss of 0.5 dB/cm, the delay-bandwidth product in Type IV is 11.1 times that in Type I. Thus, it is concluded that Type IV is preferred for group delay enhancement.

We study the nonlinear property of the system by comparing two configurations with the lowest and highest nonlinearity enhancements, i.e., Types I and IV. Figure 5(d) shows that, as loss increases, the $\chi^{(3)}$ enhancements for both types decrease. Note that the ratio of the $\chi^{(3)}$ enhancements between the Types IV and I remains almost the same, between 8.5 and 7.2. Therefore, properly choosing the operation regime of phase in the EIT-like coupled resonant systems is critically important for both group delay and nonlinearity, which can result in one order of magnitude difference.

## 3. Phase regimes of EIT-like effects in coupled *LC* resonant circuits

The phase characteristics identified above can be universally existent in various physical systems with coupled resonators. We note that the concept of the "polarity" of a resonator is applicable to various physics branches. For example, an electrical resonator with no resistance can be viewed as a lossless cavity, and changing the coupling coefficients can alter its "polarity". We find different phase responses of the EIT-like transmission profiles in electric circuits with coupled *LC* resonant structures, as an example. We numerically analyze the circuit shown in figure 6 using a circuit emulation tool. The Port A has the incident and reflected waves labeled $a_1$ and $a_2$ in figure 6, and the Port B has the incident and reflected waves named $b_1$ and $b_2$. The

S-parameter describing the response at Port A is defined as:

$$S_{11} = \frac{a_2}{a_1} \quad (5)$$

The circuit in figure 6 have two $LC$ resonators, as shown in the dash-line boxes, and their "polarities" are controlled by the in- and out-coupling coefficients determined by $L_k$ and $C_k$ ($k$ = 1, 2, 3, 4). Setting $L_1$ = 0.208 nH, $C_1$ = 4.8 pF, $L_2$ = 0.2 nH, $C_2$ = 5 pF, $L_3$ = 0.217 nH, $C_3$ = 4.6 pF, $L_4$ = 0.204 nH, $C_4$ = 4.9 pF, $L$ = 1 nH, and $C$ = 1 pF, we have the product of $L_k$ and $C_k$ ($k$ = 1, 2, 3, 4) equal to the product of $L$ and $C$. Thus, all the $LC$ resonators have the same resonance frequency. As shown in figures 7(a) and 7(b), the magnitude response of $S_{11}$ is featured by the EIT-like profile, which has a non-monotonic phase, similar to Type I in the optical resonators above. By switching $L_m$ and $C_m$ ($m$ = 1, 2) as well as $L_n$ and $C_n$ ($n$ = 3, 4), respectively, we observe the same EIT-like transmission and a $4\pi$ phase profile in figures 7(e) and 7(f). This is an analogue of the Type IV above. When we set $L_1$ = 0.195 nH, $C_1$ = 5.12 pF, $L_2$ = 0.208 nH, and $C_2$ = 4.8 pF, with the rest unchanged, a monotonic phase shift of $2\pi$ is observed, similar to the results in Types II and III. Until now, all the phase characteristics in the optical coupled resonators are also identified in the electric system.

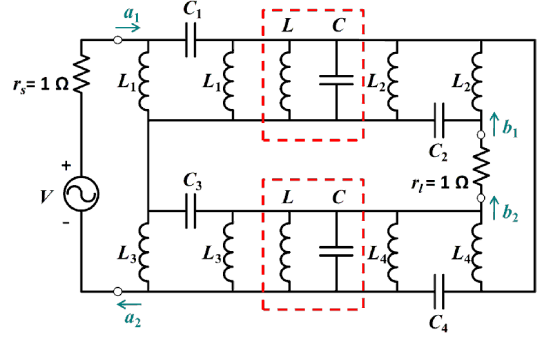

**Figure 6.** Coupled $LC$ resonant circuit exhibiting different EIT-like responses.

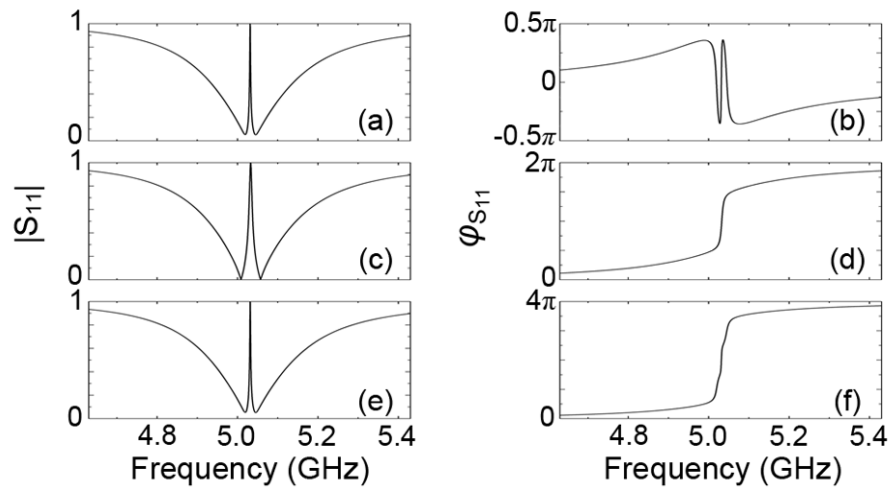

**Figure 7.** EIT-like magnitude spectra of $S_{11}$ and corresponding phase profiles. In (a) and (b), $L_1$ = 0.208 nH, $C_1$ = 4.8 pF, $L_2$ = 0.2 nH, $C_2$ = 5 pF, $L_3$ = 0.217 nH, $C_3$ = 4.6 pF, $L_4$ = 0.204 nH, $C_4$ = 4.9 pF, $L$ = 1 nH, and $C$ = 1 pF; (c) and (d) correspond to $L_1$ = 0.195 nH, $C_1$ = 5.12 pF, $L_2$ = 0.208 nH, and $C_2$ = 4.8 pF, with the rest parameters unchanged; (e) and (f) are obtained by switching $L_m$ and $C_m$ ($m$ = 1, 2) as well as $L_n$ and $C_n$ ($n$ = 3,

4) in (a) and (b), respectively.

To understand the electrical circuit in figure 6, the *LC* resonant structures enclosed in the dash-line boxes can be treated as counterparts of the optical resonators in figure 2 [45]. The sub-circuits consisting of two inductors and one capacitor on both sides of the *LC* resonators are analogous to the couplers between the optical resonators and waveguides near the resonance frequency, with the "coupling coefficients" determined by the inductance and capacitance by $\kappa_k = \mu_k^2/2 = C_k/(2CL_k)$ ($k$ = 1, 2, 3, and 4) [45]. The two electrical resonators are connected (i.e., coupled) by a wire between $L_2$ and $L_4$. In figures 7(a) and 7(b), both *LC* resonators can be regarded as under-coupled with $\kappa_1$ = 1.152×10$^{10}$ rad/s, $\kappa_2$ = 1.250×10$^{10}$ rad/s, $\kappa_3$ = 1.058×10$^{10}$ rad/s, and $\kappa_4$ = 1.201×10$^{10}$ rad/s. The upper *LC* resonator is shifted to be over-coupled in figures 7(c) and 7(d) with $\kappa_1$ = 1.311×10$^{10}$ rad/s and $\kappa_2$ = 1.152×10$^{10}$ rad/s, and the "polarity" of the lower *LC* resonator is unchanged. In figure 7(e) and 7(f), with $L_m$ and $C_m$ ($m$ = 1, 2) as well as $L_n$ and $C_n$ ($n$ = 3, 4) in (a) and (b) switched, respectively, both *LC* resonators are simultaneously switched to be over-coupled. In this way, one can expect to obtain the non-monotonic, $2\pi$, and $4\pi$ phase profiles identified in the optical systems.

**4. Application in quantum phase gate based on cavity QED**

Potentially, some important physics branches can benefit from the new phase regimes in EIT-like effects. These include, but are not limited to, (i) enhanced delay/memory and nonlinearity in all-optical [16] and optomechanical [23, 24] resonator systems, (ii) possibly new parameter space of dispersion engineering of metamaterials [46-48], and (iii) novel quantum phase gates based on cavity QED [22, 35, 36]. Here, as an example, we show how the newly identified phase regimes of EIT can be used to enhance the performance of quantum phase gate.

Cavity QED is the study of interaction between light and atom/quantum dot (QD) placed in a cavity. As a fundamental quantum logic operation, controlled-phase gate can be realized based on cavity QED with optical cavities [49]. It has been shown that the accumulated phase of cavity-reflected light is dependent on the number of photon interacting with the atom/QD, which can be controlled by a pump light [35]. The maximum phase shift is obtained in the case of atom/QD saturation, when the pump power is increased to a certain level [36]. In this situation, the EIT-like cavity-atom/QD spectrum approaches the Lorentzian shape of an empty cavity. Therefore, the maximum phase shift can be calculated as the phase difference between cavity-atom/QD and "empty cavity" operations.

An on-chip realization of the controlled quantum phase gate based on a microring-QD system is shown in figure 8, which can be generalized with the optical microring resonator replaced by any other optical cavities, such as photonic crystal cavity [35], microsphere/microdisk [50], and Fabry–Pérot cavity [34]. The side-coupled cavity-QD system imposes additional optical phase, which can be measured by interfering the affected optical wave with a reference beam, based on a Mach-Zehnder interferometer, as shown in figure 8. Signal detected at the output port is $I_{out}(\omega) = |E_{out}(\omega)|^2 = |[(1-\eta)e^{i\Delta\varphi}+\eta t(\omega)E_s(\omega)]|^2$, where $E_s(\omega)$ is the electric field of the input signal, $t(\omega)$ is the transfer function of the cavity-QD system, $\eta$ is the amplitude splitting ratio of the Y-coupler, and $\Delta\varphi$ is the phase difference between the interferometer's two arms. The phase response of the cavity-QD system, i.e., *angle*($t(\omega)$), can be obtained through fitting of detected power [35]. The detected signal spectrum changes from EIT to electromagnetically induced absorption as $\Delta\varphi$ changes from 0 to $\pi$. Here, we set $\eta$ = 0.15 and $\Delta\varphi = \pi$, to produce similar results as in [35].

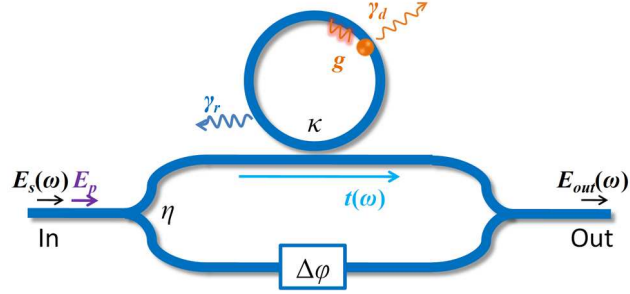

**Figure 8.** Configuration of a cavity-quantum-electrodynamics based quantum phase gate.

First, we consider the cavity-QD device operated with a signal light only. The system can be modeled with the dynamic CMT model similar to that in [22], and the transfer function $t(\omega)$ is

$$t(\omega) = \frac{[i(\omega-\omega_r)+\frac{1}{2}(\kappa-\gamma_r)][i(\omega-\omega_d)-\frac{1}{2}\gamma_d]+g^2}{[i(\omega-\omega_r)-\frac{1}{2}(\kappa+\gamma_r)][i(\omega-\omega_d)-\frac{1}{2}\gamma_d]+g^2} \quad (6)$$

where $\omega_r$ and $\omega_d$ are the resonance frequencies of the optical microring cavity and QD, respectively, $\gamma_r$ denotes the cavity decay rate due to loss, $\gamma_d$ denotes the QD spontaneous emission rate, $\kappa$ is the cavity decay rate due to coupling, and $g$ is the vacuum Rabi frequency of the QD. We note that $\gamma_d$ can be negligible compared with $\gamma_r$ [35]. Thus, the "polarity" of the cavity-QD is determined by the sign of $\kappa$-$\gamma_r$. Here, we set $\omega_r = \omega_d = \omega_0$, $\kappa/2\pi = 0.693$ GHz, $g/2\pi = 0.193$ GHz. As an example, we first consider $\gamma_r/2\pi = 1.078$ GHz ($\gamma_r > \kappa$), which is corresponding to a loss of 3.95 dB/cm (with a ring radius of 24.7 µm) in the optical cavity. In this case, $t(\omega)$ feathers an EIT-like intensity response with a phase anomaly. As shown in figures 9(a) and 9(b), the detected power and phase spectrum of $t(\omega)$ in solid lines, similar to the results in [35]. Then, when $\gamma_r/2\pi$ is decreased to 0.385 GHz ($\gamma_r < \kappa$), i.e., a loss of 1.40 dB/cm in the cavity, the operation regime is switched to the $4\pi$ phase response with a small change in transmission, as shown by solid lines in figures 9(c) and 9(d).

If a pump light is added, which saturates the QD, i.e., both $E_s$ and $E_p$ are sent into the "In" port, the transfer characteristics of the cavity-QD system can be obtained using the "empty cavity" model. The spectral responses for $\gamma_r/2\pi = 1.078$ GHz and $\gamma_r/2\pi = 0.385$ GHz are plotted with dashed lines in figures 9. A pump-light-controlled phase shift is calculated at a frequency (blue vertical line), where the maximum phase difference occurs within the phase anomaly regime, which is $0.25\pi$ in our case. This phase modulation is close to the phase shift of $0.24\pi$ demonstrated in [35]. However, the phase shift increases sharply to $0.61\pi$ in the $4\pi$ phase regime at the same frequency, which is 2.4 times that in the phase anomaly regime. If the operating frequency (blue vertical line) is blue-shifted, one can have a much larger phase difference in the $4\pi$ phase regime, but a smaller one in the phase anomaly regime. Note that a phase shift of $\pi$ can thus be achievable at $\Delta\omega = 0$, which is highly desirable in quantum computation [49] and could hardly be obtained before. The transmission at $\Delta\omega = 0$ is now around 50%, as shown in figure 9(c), which could be further increased through configuration improvement.

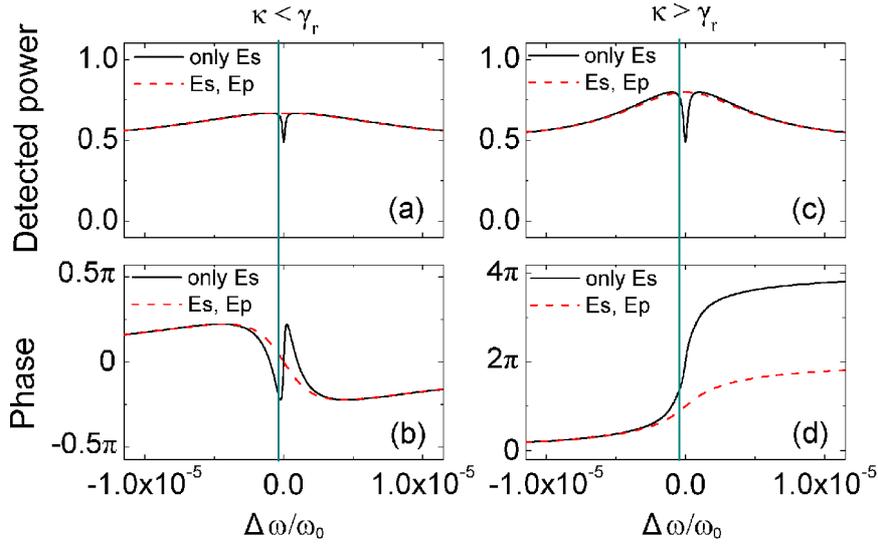

**Figure 9.** Detected power at the "Out" port and phase spectrum of the transfer function of the cavity-QD system, i.e., *angle*($t(\omega)$). (a) and (b) correspond to the case where $\kappa < \gamma_r$; (c) and (d) correspond to the case where $\kappa > \gamma_r$. Two operation conditions are considered: (i) Cavity-QD operation with only signal light ($E_s$) as input (solid line); (ii) "Empty cavity" operation with both signal light ($E_s$) and pump light ($E_p$) as inputs (dashed line).

## 5. Conclusion and Outlook

We have shown that differentiating phase regimes reveals unexplored aspects of the extensively studied EIT-like effects. One could gain a more comprehensive understanding on the operations of EIT-like devices, and thus achieve the most desirable device performance that can be dramatically different if the phase regime is switched. Very recently, it is found that coupled optical ring resonators in an embedded configuration [51] also exhibit greatly designable power and nonlinearity enhancement factors [52], which are associated with different EIT phase regimes. This serves as another good example showing the benefit of wisely choosing an EIT phase regime.

Because of the universality shown above, it is believed that the newly identified phase characteristics in the EIT-like systems can be generalized to a wide variety of coupled resonant systems. As an example, the cavity-QED-based quantum phase gate has been dramatically improved by increasing the phase tuning range up to $\pi$. Therefore, one may find other possible operation regimes in the exciting research reported in [13-26, 35, 36, 46-48] as examples, which have never been recognized and exploited.

## Acknowledgements

This paper is supported by the Natural Science Foundation of China (NSFC No.61006074 and No.61176106).